\documentstyle[epsfig,fullpage,citesort]{article}

\advance \topmargin by -\headheight
\advance \topmargin by -\headsep
     
\evensidemargin \oddsidemargin
\makeatletter

\def\@sect#1#2#3#4#5#6[#7]#8{\ifnum #2>\c@secnumdepth
     \def\@svsec{}\else 
     \refstepcounter{#1}\edef\@svsec{\csname the#1\endcsname.\hskip 1em }\fi
     \@tempskipa #5\relax
      \ifdim \@tempskipa>\z@ 
        \begingroup #6\relax
          \@hangfrom{\hskip #3\relax\@svsec}{\interlinepenalty \@M #8\par}
        \endgroup
       \csname #1mark\endcsname{#7}\addcontentsline
         {toc}{#1}{\ifnum #2>\c@secnumdepth \else
                      \protect\numberline{\csname the#1\endcsname}\fi
                    #7}\else
        \def\@svsechd{#6\hskip #3\@svsec #8\csname #1mark\endcsname
                      {#7}\addcontentsline
                           {toc}{#1}{\ifnum #2>\c@secnumdepth \else
                             \protect\numberline{\csname the#1\endcsname}\fi
                       #7}}\fi
     \@xsect{#5}}

\renewcommand{\section}{\setcounter{equation}{0} \@startsection {section}{1}
   {\z@}{-3.5ex plus -1ex minus -.2ex}{2.3ex plus .2ex}{\Large\bf}}
\newcommand{\im}{\mathop{\mathrm{Im}}}
\newcommand{\re}{\mathop{\mathrm{Re}}}

\def\IJMP #1 #2 #3 {{\it Int.\ J.\ Mod.\ Phys.}\ {\bf #1}\ (#2) #3}
\def\MPL #1 #2 #3 {{\it Mod.\ Phys.\ Lett.}\ {\bf #1}\ (#2) #3}
\def\NC #1 #2 #3 {{\it Nuovo Cim.}\ {\bf #1} (#2) #3}
\def\NP #1 #2 #3 {{\it Nucl.\ Phys.}\ {\bf #1}\ (#2) #3}
\def\PL #1 #2 #3 {{\it Phys.\ Lett.}\ {\bf #1}\ (#2) #3}
\def\PR #1 #2 #3 {{\it Phys.\ Rev.}\ {\bf #1}\ (#2) #3}
\def\PP #1 #2 #3 {{\it Phys.\ Rep.}\ {\bf #1}\ (#2) #3}
\def\PRL #1 #2 #3 {{\it Phys.\ Rev.\ Lett.}\ {\bf #1}\ (#2) #3}
\def\RMP #1 #2 #3 {{\it Rev.\ Mod.\ Phys.}\ {\bf #1}\ (#2) #3}
\def\CMP #1 #2 #3 {{\it Comm.\ Math.\ Phys.}\ {\bf #1}\ (#2) #3}
\def\ZP #1 #2 #3 {{\it Z.\ Phys.}\ {\bf #1}\ (#2) #3}
\def\E #1 #2 #3 {{\bf #1}\ (#2) #3 (E)}

\def\SM{$\cal{SM}$}
\def\sw{s_W}
\def\cw{c_W}
\def\mW{M_W}
\def\mZ{M_Z}
\def\mH{m_H}
\def\const{\frac{e^4}{(16\pi^2)^2\sw^4} \frac{\mH^2}{\mW^2}}

\begin{document}

\begin{titlepage}

\begin{flushright}
hep-ph/9712419\\
Freiburg--THEP 97/28\\
November 1997
\end{flushright}
\vspace{1.5cm}

\begin{center}
\large\bf
{\LARGE\bf Two-loop large Higgs mass corrections to electroweak gauge boson 
quartic couplings}

\vspace*{1cm}
\rm
{V. Borodulin}\\[.5cm]

{\em 	Institute for High Energy Physics}\\
{\em	Protvino, Moscow Region 142284, Russia}\\[1.5cm]
{G. Jikia}\\[.5cm]

{\em Albert--Ludwigs--Universit\"{a}t Freiburg,
           Fakult\"{a}t f\"{u}r Physik}\\
      {\em Hermann--Herder Str.3, D-79104 Freiburg, Germany}\\[1.5cm]
      
\end{center}
\normalsize

\begin{abstract}
The two-loop corrections to the electroweak gauge boson quartic
couplings, growing quadratically with the Higgs boson mass, are
calculated in the Standard Model in the limit of large Higgs mass.
The corrections to $WWWW$, $WWZZ$ and $ZZZZ$ four-vertices are found
to be an order of magnitude larger than the two-loop $m_H^2$
corrections to light fermion and triple gauge boson vertices. For a
heavy Higgs boson with a mass around 1~TeV the corrections are at the
several percent level and in principle could be observed experimentally.
\end{abstract}

\vspace{3cm}

\end{titlepage}


\section{Introduction}

The remarkable precision of the electroweak experimental data
\cite{LP'97-1,LP'97-2} makes it possible to test the predictions of
the Standard Model (\SM) at the quantum loop level. After the
successful prediction of the top-quark mass from the $m_t^2$ one-loop
electroweak radiative corrections and the actual observation of the
top quark signal at the Tevatron, the mechanism of the spontaneous
electroweak symmetry breaking, connected to the existence of the Higgs
boson in the \SM{}, remains the last untested property of the
\SM{}. Electroweak observables are influenced also by the presence of
the Higgs boson, but contrary to the $m_t^2$ dependence at the
one-loop level they depend only logarithmically on the Higgs boson
mass. From the high-precision data at LEP, SLC and the Tevatron an
upper limit of $m_H<430$~GeV has been derived at the 95\% confidence
level \cite{LP'97-1,LP'97-2}. This bound is not very sharp however. It
is known, that excluding one or two observables from the global \SM{}
fit weakens the bound significantly \cite{ALR,800}.  The reason is
that the restrictive upper bound on $m_H$ depends crucially on the
world average of the effective electroweak mixing angle $s_{eff}^2$,
for which the experimental average values from LEP and SLC differ by
2.9 standard deviations \cite{LP'97-1,LP'97-2}. As an illustration of
this situation it has been shown recently \cite{800}, that employing
only the LEP average one obtains a 95\%C.L. upper bound for the Higgs
mass larger than 800~GeV, while using the SLD value alone the
corresponding bound is approximately 80~GeV. In a conservative
conclusion the experimental limit may therefore be interpreted in the
\SM{} as an indication for a scale $m_H\leq {\cal O}(1)$~TeV.

In order to evaluate the heavy Higgs signal at high energy and
estimate the region of applicability of the perturbation theory, the
leading two--loop corrections of enhanced electroweak strength were
under intense study. In particular, the high energy weak--boson
scattering in the limit $s\gg m_H^2\gg M_W^2$ \cite{scattering},
renormalization constants at the Higgs pole
\cite{h-numeric,higgs-pole,MDR,h-analytic}, corrections to the partial
widths of the Higgs boson decay to pairs of fermions
\cite{fermi_G,fermi_DKR,h-analytic} and intermediate vector bosons
\cite{vector_G,vector_FKKR}, corrections to the heavy Higgs line shape
at LHC \cite{LHC} and $\mu^+\mu^-$ collider \cite{mumu} have been
calculated at two--loops to leading order in $m_H^2$. In addition,
recently nonperturbative next-to-leading corrections to the Higgs
propagator and Higgs boson parameters have been calculated in the
$1/N$ expansion \cite{1/N}.

However, if the Higgs boson is really heavy the study of its indirect
effects at the quantum loop level at energies much smaller than $m_H$
will be one of the most important goals for the future experiments.
As it was already mentioned, although at the one-loop level one might
expect contributions to the $W$, $Z$ vector boson mass shifts to be
proportional to $m_H^2$, for $m_H\gg M_W$ the leading terms cancel out
and only the $\log m_H^2$ dependence survives \cite{screening}.  This
has been referred to by Veltman as a screening theorem
\cite{screening}. Motivated partly by this phenomenon, van der Bij and
Veltman calculated the two-loop large Higgs mass corrections to the
$\rho$-parameter \cite{rho} and to vector boson masses \cite{masses}.
These results were verified in Ref. \cite{pisa}.  In these papers it
has been shown that, although some of the diagrams are proportional to
$m_H^4$, they cancel out in observable corrections, leaving only terms
proportional to $m_H^2$. It has been proven lately to all orders that
in the \SM{} vector propagators can contribute to low energy
observable quantities at most $(m_H^2)^{(L-1)}$ dependence on the
Higgs boson mass at the $L$-loop level for $m_H\gg M_W$
\cite{proof}. These two-loop large Higgs mass calculations were
extended by van der Bij to the case of the triple vector boson
couplings \cite{triple}. No cancellations of the leading terms happen
in this case and the two-loop corrections growing like $m_H^2$ were
found in agreement with the naive power counting arguments.  The same
power counting shows, that only vertex functions with maximally four
vector boson external legs can have two-loop large Higgs mass
corrections proportional to $m_H^2$, while for five and higher point
vertex functions no power growth of the two-loop corrections with the
Higgs mass is possible.

The main objective of the present paper is to complete these
calculations and obtain the analytical expressions for the two-loop
$m_H^2$ corrections to quartic electroweak gauge boson couplings in
the \SM{} in the limit $m_H\gg M_W$ at low energy $E\ll m_H$, and
thereby to obtain the complete two-loop low energy \SM{} effective
action $\Gamma_{eff}$ to order $m_H^2$.

The paper is organized as follows: after some preliminary discussion
of the calculational framework in Section~2, we describe the details
of the calculation in Section~3. In Section~4 all the analytical
results are presented. Section~5, where numerical results are given,
is devoted to a discussion of the implications for the physical
processes. Section~6 contains our conclusions.

\section{Calculational framework}

The calculations are done for the \SM{} in the 't~Hooft--Feynman
gauge.  We neglect fermion masses, so only bosonic loop diagrams
contribute \cite{rho}.  In order to calculate the four vector boson
vertex function contribution to the low energy effective action
$\Gamma_{eff}$ one has to take into account both one-particle
irreducible (OPI) four-vertex graphs and one particle Higgs reducible
graphs with four external vector particles, as shown in Fig.~1. Since
the Higgs self energy at two-loop (one-loop) level is proportional to
$m_H^6$ ($m_H^4$) and $HW^+W^-$, $HZZ$, $H\gamma\gamma$, $H\gamma Z$
triple vertices at two-loop (one-loop) level grow like $m_H^4$
($m_H^2$), these Higgs reducible graphs do contribute to the leading
$m_H^2$ dependence inspite of the $1/m_H^2$ suppression due to the
Higgs propagators. No one-particle reducible graphs contribute to two-
and three-point vertices and at two-loop order only quartic
$W^+W^-W^+W^-$, $W^+W^-ZZ$ and $ZZZZ$ vertices include a $m_H^2$
contribution from two-loop Higgs self energy. Due to this fact, these
vertices play a special role as a probe of the mechanism of the
electroweak symmetry breaking sector. Since $H\gamma\gamma$, $H\gamma
Z$ vertices are equal to zero at tree-level, although the Higgs
reducible graphs also contribute to the $\gamma ZW^+W^-$, $\gamma
\gamma W^+W^-$, $\gamma\gamma ZZ$ and other four-vertices with at
least one external photon, only one-loop Higgs self energy graphs
contribute.

\begin{figure}
\setlength{\unitlength}{1cm}
\begin{picture}(15,5)
\put(4,0){\epsfig{file=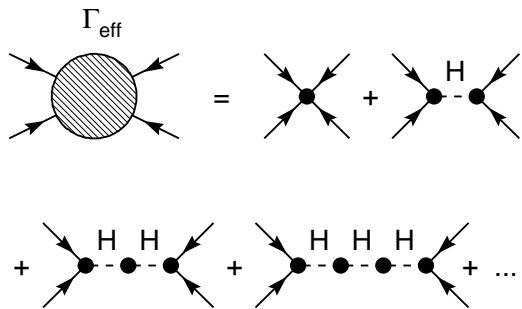,height=5cm}}
\end{picture}
\caption{One particle irreducible and Higgs reducible graphs
contributing to low energy quartic vector boson vertex. Bold blobs
denote the one particle irreducible four-, three- and two-point vertex
functions.}
\end{figure}

The calculations are done in the on--mass--shell renormalization
scheme \cite{OS}.  In this scheme all counterterms are fixed uniquely
by the requirements that the pole positions of the Higgs, $W$ and
$Z$ boson propagators coincide with their physical masses, the
corresponding residues are normalized to unity and electric charge is
renormalized to give the observable value at low energy.  

Since a heavy Higgs boson is a highly unstable particle there is an
ambiguity in the definition of its mass \cite{pole,stuart}. If the
exact Higgs propagator is defined by
\begin{equation} 
\Delta_H(s) = \frac{1}{s-m_H^2-\Sigma(s)},
\end{equation} 
one can define the Higgs boson mass either as a zero of the real part
of the inverse propagator
\begin{equation} 
\re\left(\Delta_H^{-1}(m_H^2)\right) = 0
\end{equation} 
or as a real part of the complex pole
\begin{equation} 
\Delta_H^{-1}(m_H^2-im_H\Gamma_H) = 0.
\end{equation} 
Only the complex pole mass value was shown to be gauge invariant
\cite{pole,stuart}. In our case, however, since the leading $m_H^2$
term vanishes for the derivative of the imaginary part of Higgs self
energy at the one-loop level
\begin{equation} 
\im\left(\Sigma^{1-loop}\right)'(m_H^2)=0
\end{equation} 
both definitions give the same value of the Higgs boson mass at the
two-loop order. The necessary Higgs wave function renormalization
constant and Higgs mass counterterm were calculated analytically in
our paper \cite{h-analytic} in complete agreement with the partly
numerical results \cite{h-numeric,MDR,fermi_G,fermi_DKR}.  The
two-loop wave function and mass counterterms for electroweak gauge
bosons are known since papers \cite{rho,masses}.

As was mentioned in Ref. \cite{h-analytic} there are two equivalent
definitions of the on--mass--shell renormalization scheme at two-loop
order: the standard one, when the one--loop counterterms are
calculated including terms of ${\cal O}(\epsilon)$ order, since terms
proportional to $\epsilon$ can combine with $1/\epsilon$ poles at the
two-loop level to give a finite contribution, and a modified one, when
one-loop counterterms are calculated only to ${\cal O}(\epsilon^0)$
order. These two schemes are equivalent because the account of finite
contributions coming from the combination of ${\cal O}(\epsilon)$
one--loop counterterms with $1/\epsilon$ overall divergence just
redefines the finite parts of the two--loop counterterms.  But due to
tadpole diagrams special care should be taken using the modified
scheme.  Given the part of the \SM{} Lagrangian describing the Higgs
scalar sector
\begin{eqnarray}
{\cal L} &=& \frac{1}{2}\partial_\mu H_0\partial^\mu H_0
+ \frac{1}{2}\partial_\mu z_0\partial^\mu z_0
+ \partial_\mu w^+_0\partial^\mu w^-_0 
\nonumber\\
&-&\, \frac{{m_H^2}_0}{2v_0^2}\left(w_0^+w_0^- + \frac{1}{2}z_0^2
+ \frac{1}{2}H_0^2 + v_0 H_0 + \frac{1}{2}\delta v^2
\right)^2,
\label{lagrangian}
\end{eqnarray}
one can choose the tadpole counterterm $\delta v^2$ in such a way,
that a Higgs field vacuum expectation value is equal to $v_0$
\begin{equation}
v_0 = \frac{2\, {M_W}_0}{g}
\end{equation}
to all orders, so that tadpole diagrams and corresponding counterterms
always cancel out and one can just ignore all the tadpole diagrams
altogether. If all the particles are massive it is enough to require
the cancellation of the one-loop tadpole diagrams with $\delta v^2$
counterterm to order ${\cal O}(\epsilon^0)$, because for any two-loop
Higgs reducible diagram with one-loop tadpole the one-loop subdiagram
excluding the tadpole is finite when summed with the corresponding
counterterms and terms proportional to $\epsilon$ do not
contribute. This however is only true if this subdiagram is infrared
finite, since counterterms cancel only ultraviolet divergences. If
infrared $1/\epsilon$ poles are present, they can combine with ${\cal
O}(\epsilon)$ terms resulting from incomplete tadpole and $\delta v^2$
counterterm cancellation to produce finite nonzero contribution. And
this really happens for the $W^+W^-W^+W^-$ vertex.  In order to still
validate the neglect of the tadpole diagrams one should either take
into account the one-loop tadpole $\delta v^2$ counterterm to order
${\cal O}(\epsilon)$ or regulate infrared divergences introducing
an infinitesimal photon mass $\lambda$.  Coincidence of the renormalized
vertices in both schemes was one of the consistency checks of the
calculation.

\section{The calculation}

The two-loop topologies and one-loop topologies with counterterm
insertions contributing to OPI four-, three-, and
two-point vertex functions are shown in Fig.~2. The numbers in
parentheses show the total number of corresponding topologies, the
external lines are assumed to be topologically different.

\begin{figure}
\setlength{\unitlength}{1cm}
\begin{picture}(15,10)
\put(3.5,0){\epsfig{file=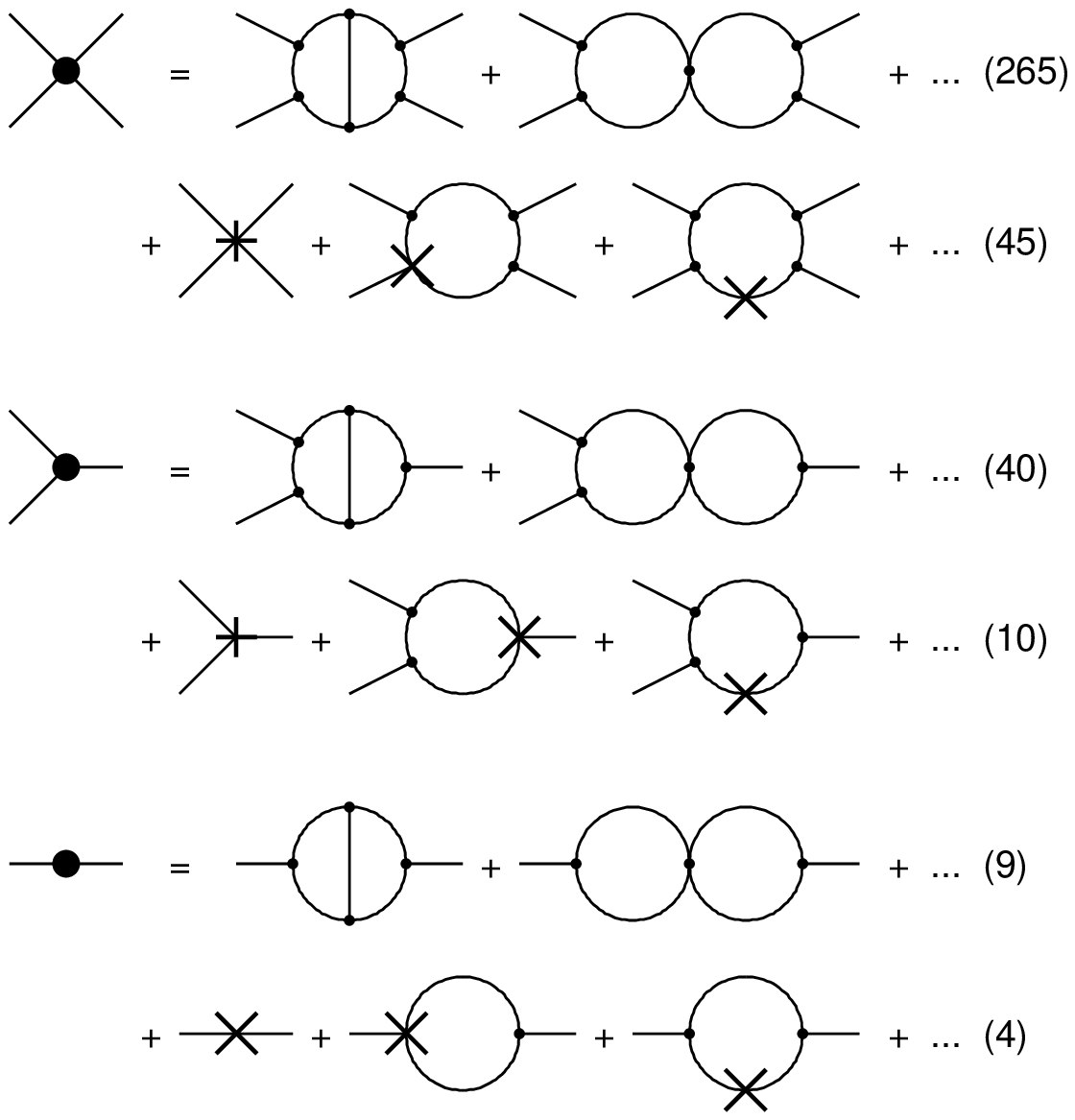,height=10cm}}
\end{picture}
\caption{One-particle irreducible two- and one-loop topologies.}
\end{figure}

The evaluation of these graphs proceeds in a number of steps.

{\it The first step} consists of the generation of all topologically
distinct graphs followed by the assignment of particles to the
internal lines according to the \SM{} Feynman rules. During this
procedure duplicated graphs are eliminated and symmetry factors are
evaluated. The total number of graphs generated is quite large. {\it
E.g.}, for irreducible $W^+W^-W^+W^-$ vertex 84698 two-loop diagrams
and 2424 one-loop diagrams contribute, but of course, not all of them
produce the $m_H^2$ dependence.

Since contributions growing with the Higgs mass at energies much
smaller than $m_H$ are described by the low energy effective action
$\Gamma_{eff}$ given by a set of local operators, containing only a
finite number of derivatives, these operators can be evaluated making
Taylor expansion around zero in the external momenta. According to
power counting only the first term of the expansion has a leading
$m_H^2$ dependence for four-point vertex functions with physical
vector bosons as external particles. For these vertices it is enough
just to set all external momenta equal to zero. However, for three-
and two-point functions and four-point functions with unphysical
Nambu-Goldstone scalar external particles one should keep more terms
in the Taylor expansion. So at the {\it second step} explicit
expressions for all the vertices are substituted and when necessary
Taylor expansions are made.  After this (trivial for four-vertices
with physical external particles) step one is left with the sum of the
two-loop vacuum diagrams.

{\it The third step} consists of the elimination of diagrams that do
not grow like a positive power of the Higgs mass. A procedure based on
the so-called asymptotic operation method \cite{Tkachov} is used to
count the maximal power of growth with $m_H$ corresponding to a given
vacuum integral.

{\it The fourth step} consists of the reduction of the remaining tensor
vacuum integrals to scalar ones, removing scalar products of loop
momenta in numerators and of the splitting of propagators with the
same momentum but different masses via partial fractioning relations.
After that everything is expressed in terms of the scalar vacuum
two-loop integrals of the following form
\begin{eqnarray}
&&J(n_1\, m_1^2,n_2\, m_2^2,n_3\, m_3^2) =\label{bubble}\\
&& -\frac{1}{\pi^4}\int\,D^{(d)}P\,D^{(d)}Q  \,\biggl(P^2-m_1^2\biggr)^{-n_1}
\biggl(Q^2-m_2^2\biggr)^{-n_2}
\biggl((P-Q)^2-m_3^2\biggr)^{-n_3}.
\nonumber
\end{eqnarray}

At {\it step five} the integrals (\ref{bubble}) are reduced using the
recurrence relations based on the integration by parts technique
\cite{parts} to the integrals of the same form (\ref{bubble}) with all
the powers $n_i\leq 1$.

And at {\it sixth step} the remaining scalar vacuum integrals are
calculated using if necessary the asymptotic expansions method
\cite{Tkachov}, if both heavy Higgs boson mass $m_H$ and light $M_W$,
$M_Z$ masses enter the integral (\ref{bubble}).

The procedure for the evaluation of the one-loop diagrams with the
counterterm insertions is essentially the same, but of course much
simpler.

Calculated OPI graphs are then substituted into the Higgs reducible
graphs, as shown in Fig.~1, to obtain corresponding operators of the
low energy effective action $\Gamma_{eff}$. As a result all the
amplitudes are expressed in terms of transcendental functions
$\zeta(3)$ and the maximal value of the Clausen function
$\mbox{Cl}(\pi/3)$.

In order to check the results obtained in addition to obvious
consistency checks like the finiteness of the renormalized results and
symmetry properties, the other cross checks were also done.  As was
mentioned at the end of Section~2, we have done the calculations in
two different renormalization schemes: one with zero photon mass and
one-loop counterterms calculated to order ${\cal O}(\epsilon)$ and the
other with small nonzero photon mass $\lambda$ and one-loop
counterterms calculated to order ${\cal O}(\epsilon^0)$. The results
for the renormalized effective action were found to be
identical. Moreover, some individual diagrams are infrared divergent
and since there are no tree-level or one-loop vertices with the
emission of additional photon, which would grow like $m_H^2$, this
infrared divergences should be canceled out in the sum of all
contributing diagrams. In the second renormalization scheme it is
possible to explicitly observe that all singular logarithms
$\log\lambda^2$ really cancel out.

As a last verification of the calculation the Ward identities in the
form, which is a basis of the equivalence theorem \cite{ET},
\begin{eqnarray}
&&\langle 0|T\left[\biggl(\partial^\mu W^+_\mu(x_1) +\mW w^+(x_1)\biggr)
W^-_\nu(x_2)W^+_\alpha(x_3)W^-_\beta(x_4)\right]|0\rangle_{amp}=0,\label{ET}\\
&&\langle 0|T\left[\biggl(\partial^\mu Z_\mu(x_1) +\mZ z(x_1)\biggr)
Z_\nu(x_2)W^+_\alpha(x_3)W^-_\beta(x_4)\right]|0\rangle_{amp}=0,\quad
\dots\nonumber
\end{eqnarray}
have been checked for all the four-particle amplitudes. Here $w$ and
$z$ are the unphysical Nambu-Goldstone partners of $W$, $Z$ bosons.
In order to check the identities (\ref{ET}) all the amplitudes with
three external vector bosons and one non-physical scalar were
calculated.  If the first external vector boson is a photon, then the
identity (\ref{ET}) is just a transversality condition. The
chronological product here is in fact a complete renormalized
four-particle Feynman amplitude with amputated external legs,
including OPI as well as one-particle reducible graphs with light
particle propagators. Usually the Ward identities (\ref{ET}) are
formulated for unrenormalized fields and non-amputated external leg
for $\partial^\mu W$, $w$ ($\partial^\mu Z$, $z$) lines \cite{ET},
however taking into account that non-diagonal self-energies
(\ref{wW}), (\ref{zZ}) also satisfy corresponding Ward identities
\begin{eqnarray}
i k^\mu \langle 0|T\left[W^+_\mu(k) W^-_\nu(-k)\right]|0\rangle&=&
\mW \langle 0|T\left[w^+(k)W^-_\nu(-k)\right]|0\rangle,\\
i k^\mu \langle 0|T\left[Z_\mu(k) Z_\nu(-k)\right]|0\rangle&=&
\mZ \langle 0|T\left[z(k) Z_\nu(-k)\right]|0\rangle,
\nonumber
\end{eqnarray}
one can show that the identities (\ref{ET}) are really valid to
two-loop order.

\section{Analytic results}

Here we present all the terms of the two-loop low-energy effective
action $\Gamma_{eff}$ to order $m_H^2$, which are necessary for the
calculation of the electroweak gauge boson scattering amplitudes and
amplitudes for the reactions of triple $WWZ/ZZZ$ production at the
CERN Large Hadron Collider (LHC) or the electron-positron linear
collider, which are sensitive to quartic vector boson couplings:
\begin{eqnarray}
&&ee\to VVff,\label{ee-nnVV},\\
&&ee\to VVV\label{ee-VVV},\\
&&pp\to VVX,\label{LHC},\\
&&pp\to VVVX,
\end{eqnarray}
where $V=\gamma$, $Z$ or $W^\pm$ and $f=e$ or $\nu_e$.

The fermionic part of the effective action is defined by
\begin{eqnarray}
\Gamma^{fermi}_{eff} &=& eQ\bar{f}\hat{A}f
\,+\, \frac{g_{eff}}{\sqrt{2}}\left(\bar{f}_L\hat{W}^+f'_L
\,+\, \bar{f'}_L\hat{W}^-f_L\right)\nonumber\\
&&+\, \bar{g}_{eff}\left(\bar{f}_L(T_3-Q{\sw^2}_{eff})\hat{Z}f_L
-\bar{f}_R Q{\sw^2}_{eff}\hat{Z}f_R\right).
\end{eqnarray}
Here effective coupling constants are observable quantities, that are
measured in the process of $\mu$ decay, in fermion scattering
reactions at the $Z$ boson peak, {\it etc}. They are given by the
following expressions
\begin{eqnarray}
g_{eff} &=& \frac{e}{\sw}\left\{1 + \const\left(
       -  \frac{11}{64} \pi  \sqrt{3} - \frac{25}{1728}\pi^2   
       + \frac{49}{1152} + \frac{9}{16} Cl \sqrt{3}\right)\right\}\nonumber\\
&\approx&\frac{e}{\sw}(1  -4.66591\times 10^{-2} \const)\,\approx\,
\frac{e}{\sw}(1  -3.15948\times 10^{-7} \frac{\mH^2}{\mW^2}),
\label{g}\\
\bar{g}_{eff} &=& \frac{e}{\sw\cw}\biggl\{1 + \const\biggl(
       \frac{1}{\cw^2}   \biggl( \frac{9}{64} \pi   \sqrt{3}
       +   \frac{3}{64} \pi^2 
       - \frac{21}{128} - \frac{9}{16} Cl \sqrt{3} \biggr)\nonumber\\
&&       - \frac{5}{16} \pi   \sqrt{3}
       -  \frac{53}{864} \pi^2  
       + \frac{119}{576} + \frac{9}{8} Cl \sqrt{3}\biggr)\biggr\}
\label{gbar}\\
&\approx&\frac{e}{\sw\cw}(1 -2.51326\times 10^{-2} \const)\,\approx\,
\frac{e}{\sw\cw}(1 -1.70183\times 10^{-7} \frac{\mH^2}{\mW^2}),
\nonumber\\
{\sw^2}_{eff} &=& \sw^2\left\{1 + \const\left(
       +   \frac{5}{16} \pi \sqrt{3}
       +  \frac{53}{864} \pi^2  
       - \frac{119}{576} - \frac{9}{8} Cl \sqrt{3}\right)\right\}\nonumber\\
&\approx&\sw^2(1 + 0.121595\, \const)\,\approx\,
\sw^2(1 +8.23369\times 10^{-7} \frac{\mH^2}{\mW^2}).\label{sw}
\end{eqnarray}
Here 
\begin{equation}
Cl = \mbox{Cl}(\frac{\pi}{3}) = \im \,
\mbox{li}_2(e^{\frac{i\pi}{3}})= 
1.01494\: 16064\: 09653\: 62502\dots
\end{equation}
and $\cw$ in the on-mass-shell scheme is defined by
\begin{equation}
\cw = \frac{\mW}{\mZ}.
\end{equation}
As was demonstrated in Ref. \cite{rho}, in the limit of vanishing
fermion mass, no $m_H^2$ contributions appear for two-loop diagrams
with external fermion lines with subtracted one-loop subdivergences.
Thus, the corrections (\ref{g})-(\ref{sw}) originate only from vector
boson wave function, $\sw$, $\cw$ and electric charge renormalizations
at two-loop level. The expressions (\ref{g})-(\ref{sw}) are equivalent
to the results of \cite{rho,pisa} and for the observable quantities
like $\delta\rho$ give the identical values.

The renormalized self-energies of light particles are given by
\begin{eqnarray}
\frac{\delta\Gamma_{eff}}{\delta W^+_{\mu_1}(k_1)\delta W^-_{\mu_2}(k_2)} &=&
        \frac{e^4}{(16\pi^2)^2\sw^4} \frac{\mH^2}{\mW^2} 
	k_1^{\mu_1} k_1^{\mu_2}  \Biggl\{
          - \frac{71}{192} 
          + \frac{11}{288}  \pi^2
          \Biggr\},
\label{WW}\\
\frac{\delta\Gamma_{eff}}{\delta Z_{\mu_1}(k_1)\delta Z_{\mu_2}(k_2)}  &=&
        \frac{e^4}{(16\pi^2)^2\cw^2\sw^4} \frac{\mH^2}{\mW^2}   
	k_1^{\mu_1} k_1^{\mu_2} \Biggl\{
          - \frac{71}{192} 
          + \frac{11}{288} \pi^2
          \Biggr\},
\label{ZZ}\\
\frac{\delta\Gamma_{eff}}{\delta w^+(k_1)\delta w^-(k_2)}&=&
        \frac{e^4}{(16\pi^2)^2\sw^4} \frac{\mH^2}{\mW^4}   
	(k_1\cdot k_1)^2 \Biggl\{
          - \frac{71}{192} 
          + \frac{11}{288}  \pi^2
          \Biggr\},
\label{ww}\\
\frac{\delta\Gamma_{eff}}{\delta z(k_1)\delta z(k_2)}  &=&
        \frac{e^4}{(16\pi^2)^2\sw^4} \frac{\mH^2}{\mW^4}   
	(k_1\cdot k_1)^2 \Biggl\{
          - \frac{71}{192} 
          + \frac{11}{288} \pi^2
          \Biggr\},
\label{zz}\\
\frac{\delta\Gamma_{eff}}{\delta w^+(k_1)\delta W^-_{\mu_2}(k_2)} &=&
        i \frac{e^4}{(16\pi^2)^2\sw^4} \frac{\mH^2}{\mW^3}   
	k_1^{\mu_2} k_1\cdot k_1 \Biggl\{
          - \frac{71}{192} 
          + \frac{11}{288} \pi^2
          \Biggr\},
\label{wW}\\
\frac{\delta\Gamma_{eff}}{\delta z(k_1)\delta Z_{\mu_2}(k_2)} &=&
        i \,\frac{e^4}{(16\pi^2)^2\sw^4\cw} \frac{\mH^2}{\mW^3}   
	k_1^{\mu_2} k_1\cdot k_1 \Biggl\{
          - \frac{71}{192} 
          + \frac{11}{288} \pi^2
          \Biggr\},
\label{zZ}\\
\frac{\delta\Gamma_{eff}}{\delta A_{\mu_1}(k_1)\delta A_{\mu_2}(k_2)} &=&
\frac{\delta\Gamma_{eff}}{\delta A_{\mu_1}(k_1)\delta Z_{\mu_2}(k_2)} =
\frac{\delta\Gamma_{eff}}{\delta z(k_1)\delta A_{\mu_2}(k_2)} = 0.
\end{eqnarray}
Since mass and residue of the transverse part of the vector boson
propagators are fixed via the renormalization conditions, only the
finite longitudinal structure survives in (\ref{WW}), (\ref{ZZ}).

Triple vertices with external physical electroweak gauge bosons and
triple vertices with one external unphysical scalar, which are also
needed for the calculation of the vector boson scattering amplitudes
in 't~Hooft--Feynman gauge, are given by
\begin{eqnarray}
&& \frac{\delta\Gamma_{eff}}
{\delta A_{\mu_1}(k_1)\delta W_{\mu_2}^+(k_2)\delta W_{\mu_3}^-(k_3)} = 
\frac{e^5}{(16\pi^2)^2\sw^4}\,\frac{\mH^2}{\mW^2}   \Biggl\{
+ \left(g^{\mu_1\mu_2} k_2^{\mu_3} - g^{\mu_1\mu_3} k_3^{\mu_2}\right) 
	  \left(  - \frac{7}{12} + \frac{1}{18} \pi^2 \right)
\nonumber\\
&& + \left(g^{\mu_1\mu_2} k_3^{\mu_3} - g^{\mu_1\mu_3} k_2^{\mu_2}\right) 
     \left(  - \frac{41}{192} + \frac{5}{288} \pi^2 \right)
		                 \Biggr\} 
\label{AWW}\\
&&\approx\frac{e^5}{(16\pi^2)^2\sw^4}\,\frac{\mH^2}{\mW^2}   \Biggl\{
  -3.50220\times 10^{-2}\,
\left(g^{\mu_1\mu_2} k_2^{\mu_3} - g^{\mu_1\mu_3} k_3^{\mu_2}\right) 
 -4.21944\times 10^{-2}\,
\left(g^{\mu_1\mu_2} k_3^{\mu_3} - g^{\mu_1\mu_3} k_2^{\mu_2}\right) 
		                 \Biggr\},
\nonumber
\\[4mm]
&&\frac{\delta\Gamma_{eff}}
{\delta Z_{\mu_1}(k_1)\delta W_{\mu_2}^+(k_2)\delta W_{\mu_3}^-(k_3)}  = 
-\frac{\sw}{\cw}\frac{\delta\Gamma_{eff}}
{\delta A_{\mu_1}(k_1)\delta W_{\mu_2}^+(k_2)\delta W_{\mu_3}^-(k_3)} 
\nonumber\\
&&+\, \frac{e^5}{(16\pi^2)^2\cw\sw^5}\, \frac{\mH^2}{\mW^2}   \biggl(
   g^{\mu_1\mu_2} (k_2^{\mu_3} - k_1^{\mu_3}) 
   + g^{\mu_1\mu_3} (k_1^{\mu_2} - k_3^{\mu_2})
   + g^{\mu_2\mu_3} (k_3^{\mu_1} - k_2^{\mu_1})  \biggr)
\label{ZWW}\\
&&  \times\left(  - \frac{217}{1152} - \frac{5}{32} \pi \sqrt{3} 
- \frac{5}{1728} \pi^2 + \frac{9}{16} Cl \sqrt{3} \right)\,\approx\,
-\frac{\sw}{\cw}\frac{\delta\Gamma_{eff}}
{\delta A_{\mu_1}(k_1)\delta W_{\mu_2}^+(k_2)\delta W_{\mu_3}^-(k_3)} 
\nonumber\\
&&-\, 7.83085\times 10^{-2}\,
\frac{e^5}{(16\pi^2)^2\cw\sw^5}\, \frac{\mH^2}{\mW^2}   \biggl(
   g^{\mu_1\mu_2} (k_2^{\mu_3} - k_1^{\mu_3}) 
   + g^{\mu_1\mu_3} (k_1^{\mu_2} - k_3^{\mu_2})
   + g^{\mu_2\mu_3} (k_3^{\mu_1} - k_2^{\mu_1})  \biggr),
\nonumber
\\[4mm]
&&\frac{\delta\Gamma_{eff}}
{\delta z(k_1)\delta W^+_{\mu_2}(k_2)\delta W^-_{\mu_3}(k_3)}  = 
i \,\frac{e^5}{(16\pi^2)^2\sw^5}\frac{\mH^2}{\mW^3} \Biggl\{
+ \left(k_2^{\mu_2} k_2^{\mu_3}-k_3^{\mu_2} k_3^{\mu_3}\right) 
\left(  - \frac{5}{36} - \frac{1}{64} \pi \sqrt{3} 
+ \frac{23}{864} \pi^2 \right)
\\
&&       + g^{\mu_2 \mu_3} \left(k_2\cdot k_2-k_3\cdot k_3\right) 
\left(  - \frac{133}{576} + \frac{1}{64} \pi \sqrt{3} 
+ \frac{5}{432} \pi^2 \right)\Biggr\},
\nonumber
\\[4mm]
&&\frac{\delta\Gamma_{eff}}
{\delta w^+(k_1)\delta W^-_{\mu_2}(k_2)\delta A_{\mu_3}(k_3)}  = 
	i \,\frac{e^5}{(16\pi^2)^2\sw^4}\frac{\mH^2}{\mW^3}   \Biggl\{
+ k_2^{\mu_2} (k_2^{\mu_3} - k_1^{\mu_3}) 
\left(- \frac{71}{192} + \frac{11}{288} \pi^2 \right)
\nonumber\\
&& + k_2^{\mu_3} k_3^{\mu_2}   
\left(  - \frac{7}{12} + \frac{1}{18} \pi^2 \right)
+ k_3^{\mu_2} k_3^{\mu_3} 
\left( - \frac{133}{288} + \frac{1}{32} \pi \sqrt{3} 
	+ \frac{5}{216} \pi^2 \right)
\\
&&+ g^{\mu_2 \mu_3} k_2\cdot k_2   
\left(  - \frac{7}{24} + \frac{1}{36} \pi^2 \right)
+ g^{\mu_2 \mu_3} k_1\cdot k_1   
\left(  - \frac{5}{64} + \frac{1}{96} \pi^2 \right)
\nonumber\\
&&+ g^{\mu_2 \mu_3} k_3\cdot k_3   
\left( \frac{49}{288} - \frac{1}{32} \pi \sqrt{3} 
	+ \frac{1}{216} \pi^2 \right)
                              \Biggr\},    
\nonumber
\\[4mm]
&&\frac{\delta\Gamma_{eff}}
{\delta w^+(k_1)\delta W^-_{\mu_2}(k_2)\delta Z_{\mu_3}(k_3)} = 
-\frac{\sw}{\cw} \frac{\delta\Gamma_{eff}}
{\delta w^+(k_1)\delta W^-_{\mu_2}(k_2)\delta A_{\mu_3}(k_3)}
+\frac{1}{\cw} \frac{\delta\Gamma_{eff}}
{\delta z(k_1)\delta W^+_{\mu_2}(k_2)\delta W^-_{\mu_3}(k_3)}
\nonumber\\
&&	+ i \,\frac{e^5}{(16\pi^2)^2\cw^3\sw^3} \frac{\mH^2}{\mW}  
g^{\mu_2 \mu_3} \left( \frac{9}{64} \pi \sqrt{3} + \frac{3}{64} \pi^2 
         - \frac{9}{16} Cl \sqrt{3} - \frac{21}{128} \right).
\end{eqnarray}
As one can see from (\ref{AWW}), the electric charge of the $W^\pm$
boson is not renormalized and only the anomalous magnetic moment and
the longitudinal tensor, which is zero if both $W$ bosons are
physical, appear.  For the $ZW^+W^-$ vertex both coupling finite
constant renormalization and anomalous magnetic moment interaction
terms are present. The results for the $\gamma W^+W^-$, $ZW^+W^-$
vertices are equivalent to the results of Ref. \cite{triple} and give
the same results for the anomalous magnetic moments and corrections to
the reaction $e^+e^-\to W^+W^-$.

And finally the following quartic electroweak vector boson vertices
were obtained:
\begin{eqnarray}
&&\frac{\delta\Gamma_{eff}}{\delta W^+_{\mu_1}(k_1)\delta W^-_{\mu_2}(k_2)
\delta W^+_{\mu_3}(k_3)\delta W^-_{\mu_4}(k_4)}  =
  \,\frac{e^6}{(16\pi^2)^2\sw^6} \frac{\mH^2}{\mW^2} \Biggl\{
 \biggl(g^{\mu_1 \mu_2} g^{\mu_3 \mu_4} 
+ g^{\mu_1 \mu_4} g^{\mu_2 \mu_3}\biggr)   
\nonumber\\
&& \times\left(
   + \frac{4027}{768}
   - \frac{39}{16} \pi Cl
   + \frac{127}{64} \pi \sqrt{3}
   - \frac{3319}{3456} \pi^2
   - \frac{289}{96} Cl \sqrt{3}
   + \frac{63}{16} \zeta(3)
   \right)
\label{WWWW}\\
&&+ g^{\mu_1 \mu_3} g^{\mu_2 \mu_4}   \left(
   + \frac{17}{384}
   - \frac{11}{16} \pi \sqrt{3}
   - \frac{23}{1728} \pi^2
   + \frac{97}{48} Cl \sqrt{3}
   \right)
  \Biggr\}
\nonumber\\
&&\,\approx\,\,\frac{e^6}{(16\pi^2)^2\sw^6} \frac{\mH^2}{\mW^2} \Biggl\{
 -\,1.76815\, \biggl(g^{\mu_1 \mu_2} g^{\mu_3 \mu_4} 
+ g^{\mu_1 \mu_4} g^{\mu_2 \mu_3}\biggr)   
  -\,0.275572\, g^{\mu_1 \mu_3} g^{\mu_2 \mu_4}   \Biggr\},
\nonumber
\\[4mm]
&&\frac{\delta\Gamma_{eff}}{\delta W^+_{\mu_1}(k_1)\delta W^-_{\mu_2}(k_2)
\delta Z_{\mu_3}(k_3)\delta Z_{\mu_4}(k_4)}  =  
\,\frac{e^6}{(16\pi^2)^2\sw^6\cw^2} \frac{\mH^2}{\mW^2} \Biggl\{
\nonumber\\
&&+ g^{\mu_1 \mu_2} g^{\mu_3 \mu_4}   \Biggl[
   + \frac{5147}{1152}
   - \frac{39}{16} \pi Cl
   + \frac{109}{64} \pi \sqrt{3}
   - \frac{139}{144} \pi^2
   - \frac{85}{48} Cl \sqrt{3}
   + \frac{63}{16} \zeta(3)
\nonumber\\
&&+ \cw^2   \left(
   + \frac{217}{288}
   + \frac{5}{8} \pi \sqrt{3}
   + \frac{5}{432} \pi^2
   - \frac{9}{4} Cl \sqrt{3}
   \right)\Biggr]
\label{WWZZ}\\
&&+ \biggl(g^{\mu_1 \mu_3} g^{\mu_2 \mu_4} 
+ g^{\mu_1 \mu_4} g^{\mu_2 \mu_3}\biggr)   
\Biggl[
   + \frac{67}{2304}
   - \frac{1}{32} \pi \sqrt{3}
   + \frac{43}{1152} \pi^2
   - \frac{11}{96} Cl \sqrt{3}
\nonumber\\
&&+ \cw^2   \left(
   + \frac{209}{576}
   - \frac{5}{16} \pi \sqrt{3}
   - \frac{71}{864} \pi^2
   + \frac{9}{8} Cl \sqrt{3}
   \right)
+ \cw^4   \left(
   - \frac{71}{192}
   + \frac{11}{288} \pi^2
   \right)\Biggr]
 \Biggr\}
\nonumber\\
&&\,\approx\,\,\frac{e^6}{(16\pi^2)^2\sw^6\cw^2} \frac{\mH^2}{\mW^2} \Biggl\{
+ g^{\mu_1 \mu_2} g^{\mu_3 \mu_4}   \biggl[
	 -\,1.94360  \,+\, 0.313234\,\cw^2   \biggr]
\nonumber\\
&&+ \biggl(g^{\mu_1 \mu_3} g^{\mu_2 \mu_4} 
+ g^{\mu_1 \mu_4} g^{\mu_2 \mu_3}\biggr)   
\biggl[2.60033\times 10^{-2} 
  -\,0.170962\,\cw^2   + 7.17239\times 10^{-3}\cw^4   \biggr]
 \Biggr\},
\nonumber
\\[4mm]
&&\frac{\delta\Gamma_{eff}}{\delta Z_{\mu_1}(k_1)\delta Z_{\mu_2}(k_2)
\delta Z_{\mu_3}(k_3)\delta Z_{\mu_4}(k_4)} =  
\nonumber\\
&&\frac{e^6}{(16\pi^2)^2\sw^6\cw^4} \frac{\mH^2}{\mW^2} 
\biggl(g^{\mu_1 \mu_2} g^{\mu_3 \mu_4} + g^{\mu_1 \mu_3} g^{\mu_2 \mu_4} 
  + g^{\mu_1 \mu_4} g^{\mu_2 \mu_3}\biggr)
\label{ZZZZ}\\
&&\times \left(
          + \frac{337}{64}
          - \frac{39}{16} \pi Cl
          + \frac{105}{64} \pi \sqrt{3}
          - \frac{557}{576} \pi^2
          - 2 Cl \sqrt{3}
          + \frac{63}{16} \zeta(3)
          \right)
\nonumber\\
&&\,\approx\,-\,1.90594\,\frac{e^6}{(16\pi^2)^2\sw^6\cw^4} \frac{\mH^2}{\mW^2} 
\biggl(g^{\mu_1 \mu_2} g^{\mu_3 \mu_4} + g^{\mu_1 \mu_3} g^{\mu_2 \mu_4} 
  + g^{\mu_1 \mu_4} g^{\mu_2 \mu_3}\biggr)
\nonumber
\\[4mm]
&&\frac{\delta\Gamma_{eff}}{\delta A_{\mu_1}(k_1)\delta A_{\mu_2}(k_2)
\delta W^+_{\mu_3}(k_3)\delta W^-_{\mu_4}(k_4)}  =  
\nonumber\\
&&\frac{e^6}{(16\pi^2)^2\sw^4} \frac{\mH^2}{\mW^2} 
   \biggl(g^{\mu_1 \mu_3} g^{\mu_2 \mu_4} 
+ g^{\mu_1 \mu_4} g^{\mu_2 \mu_3}\biggr)   
\left(     - \frac{71}{192} + \frac{11}{288} \pi^2 \right)
\label{AAWW}\\
&&\,\approx\,
7.17239\times 10^{-3}\,\frac{e^6}{(16\pi^2)^2\sw^4} \frac{\mH^2}{\mW^2} 
   \biggl(g^{\mu_1 \mu_3} g^{\mu_2 \mu_4} 
+ g^{\mu_1 \mu_4} g^{\mu_2 \mu_3}\biggr),   
\nonumber
\\[4mm]
&&\frac{\delta\Gamma_{eff}}{\delta A_{\mu_1}(k_1)\delta Z_{\mu_2}(k_2)
\delta W^+_{\mu_3}(k_3)\delta W^-_{\mu_4}(k_4)}  =  
\nonumber\\
&&\frac{e^6}{(16\pi^2)^2\sw^5\cw} \frac{\mH^2}{\mW^2} \Biggl\{
+ g^{\mu_1 \mu_2} g^{\mu_3 \mu_4}   \left(
   + \frac{217}{576}
   + \frac{5}{16} \pi \sqrt{3}
   + \frac{5}{864} \pi^2
   - \frac{9}{8} Cl \sqrt{3}
   \right)
\label{AZWW}\\
&&+ \biggl(g^{\mu_1 \mu_3} g^{\mu_2 \mu_4} 
+ g^{\mu_1 \mu_4} g^{\mu_2 \mu_3}\biggr)   \Biggl[
   + \frac{209}{1152}
   - \frac{5}{32} \pi \sqrt{3}
   - \frac{71}{1728} \pi^2
   + \frac{9}{16} Cl \sqrt{3}
+ \cw^2   \left(
   - \frac{71}{192}
   + \frac{11}{288} \pi^2
   \right)\Biggr]
 \Biggr\}
\nonumber\\
&&\,\approx\,\frac{e^6}{(16\pi^2)^2\sw^5\cw} \frac{\mH^2}{\mW^2} \Biggl\{
+\, 0.156617\,g^{\mu_1 \mu_2} g^{\mu_3 \mu_4}   
\nonumber\\
&&+ \biggl(g^{\mu_1 \mu_3} g^{\mu_2 \mu_4} 
+ g^{\mu_1 \mu_4} g^{\mu_2 \mu_3}\biggr)   \biggl[
    -\,8.54809\times 10^{-2}+ 7.17239\times 10^{-3}\,\cw^2 \biggr]
 \Biggr\},
\nonumber
\\[4mm]
&&\frac{\delta\Gamma_{eff}}{\delta A_{\mu_1}(k_1)\delta A_{\mu_2}(k_2)
\delta A_{\mu_3}(k_3)\delta A_{\mu_4}(k_4)} =  
\frac{\delta\Gamma_{eff}}{\delta A_{\mu_1}(k_1)\delta A_{\mu_2}(k_2)
\delta Z_{\mu_3}(k_3)\delta Z_{\mu_4}(k_4)} = \dots = 0. 
\label{AAAA}
\end{eqnarray}
There is a $ZZZZ$ vertex, which is not present in the \SM{} at the
tree-level.  All the other vertices with neutral gauge bosons and at
least one photon vanish.  Also the $\gamma\gamma W^+W^-$ quartic
coupling is generated. Obviously, the tensor structures proportional
to $(- 71/192 + \pi^2 11/288)$, explicitly appearing in 
Eq. (\ref{AAWW}) and the expressions for the two-point functions
(\ref{WW})-(\ref{zZ}), are different terms entering the chiral
Lagrangian ${\cal L}_{11}$ \cite{chiral}
\begin{equation}
{\cal L}_{11} = Tr\left[{\cal D}_\mu{\cal V}^\mu 
{\cal D}_\nu{\cal V}^\nu \right],
\end{equation}
where the Nambu-Goldstone bosons $\omega^i$ are assembled in a unitary
matrix $\Sigma=\exp(i\omega^i\tau^i/v)$ and
\begin{equation}
{\cal D}_\mu\Sigma=\partial_\mu\Sigma+\frac{i}{2}
\left(g\hat W_\mu\Sigma-g'\Sigma\tau_3 B_\mu\right);\quad
{\cal V}_\mu = ({\cal D}_\mu\Sigma)\Sigma^\dagger.
\end{equation}

As one can see from (\ref{g})-(\ref{sw}), (\ref{AWW})-(\ref{ZWW}) the
two-loop $m_H^2$ corrections to fermion scattering processes and
triple vector boson couplings are very small, inspite of the
$m_H^2/\mW^2$ enhancements, not only because of the small two-loop
factor $g^4/(16\pi^2)^2$, but also because the dimensionless
coefficients themselves are quite small. {\it E.g.}, the largest
coefficient that enters the expression for ${\sw^2}_{eff}$ (\ref{sw})
is approximately 0.1.  The typical values for other dimensionless
coefficients are several units times $10^{-2}$. In this respect the
$W^+W^-W^+W^-$, $W^+W^-ZZ$ and $ZZZZ$ quartic couplings represent a
drastic contrast to the other vertices.  The dimensionless
coefficients in (\ref{WWWW}), (\ref{WWZZ}), (\ref{ZZZZ}) are about 2,
{\it i.e.} about 20 times larger, than for ${\sw^2}_{eff}$ (\ref{sw})!
As was mentioned previously, these particular vertices are
distinguished, due to a contribution from two-loop Higgs self energy
insertion in the Higgs-reducible graphs.  These vertices receive
a contribution from the $\zeta(3)$ and $\pi Cl$ terms, which originate
only from the two-loop Higgs mass counterterm \cite{h-analytic} as a
term proportional to a linear combination $21\zeta(3) -13\pi Cl$. In a
sense these couplings could be considered ``genuine'' quartic
couplings, which are the most sensitive to the details of the
mechanism of the electroweak symmetry breaking.

\section{Numerical results}

The possibilities to probe the quartic vector boson couplings through
the $WW$-, $ZZ$-fusion reactions (\ref{ee-nnVV}) \cite{ee-nnVV} and
through $WWZ/ZZZ$ triple gauge boson production (\ref{ee-VVV})
\cite{ee-VVV} at high energy linear colliders are under intense
study. In this Section we will not give any extensive phenomenological
analyses, but present cross-sections and distributions for the
fundamental subprocesses of electroweak gauge boson scattering reactions:
\begin{equation}
VV\to VV.\label{VVVV}
\end{equation}

In Fig.~3 we show angular distributions for processes of type
(\ref{VVVV}). We present these distributions both for unpolarized and
longitudinal vector bosons. The energy is taken to be
$\sqrt{s_{VV}}=500$~GeV and the Higgs mass $m_H=900$~GeV. We give the
results both for the tree-level and for the tree level plus the
two-loop $m_H^2$ corrections.  As one would expect, the largest
relative corrections are seen for the $ZZ\to ZZ$ reaction, which
receives only a small contribution from heavy Higgs boson exchange at
the tree level.

\begin{figure}[p]
\setlength{\unitlength}{1cm}
\begin{picture}(14,20)
\put(2,0){\epsfig{file=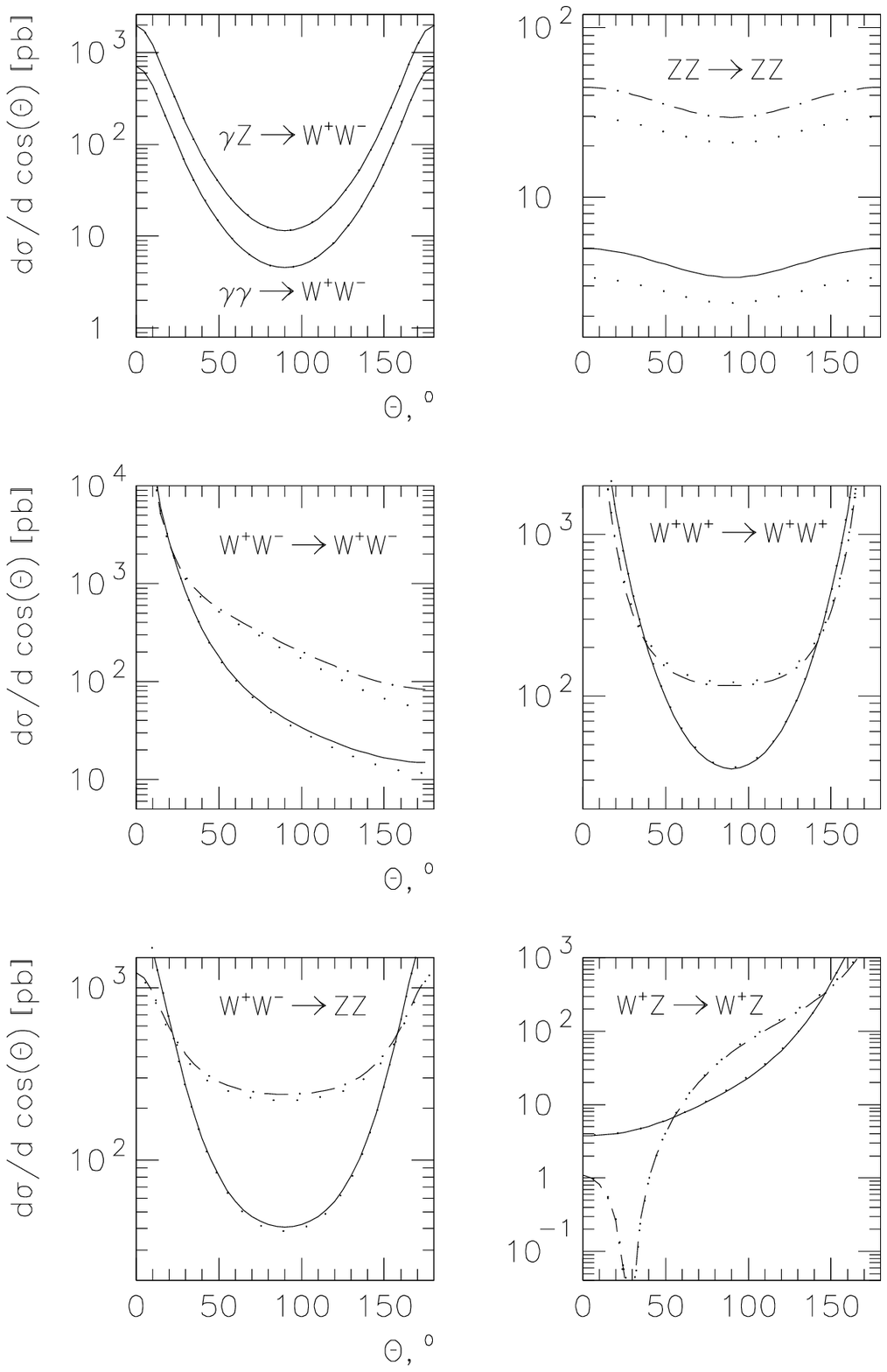,height=20cm}}
\end{picture}
\caption{Born differential cross sections for unpolarized $UUUU$
(solid line) and longitudinal $LLLL$ (dot-dashed line) vector bosons
at $\sqrt{s_{VV}}=500$~GeV and $m_H=900$~GeV. Dotted lines lines show
corresponding two-loop corrected cross sections.}
\end{figure}

\begin{figure}[p]
\setlength{\unitlength}{1cm}
\begin{picture}(14,20)
\put(2,0){\epsfig{file=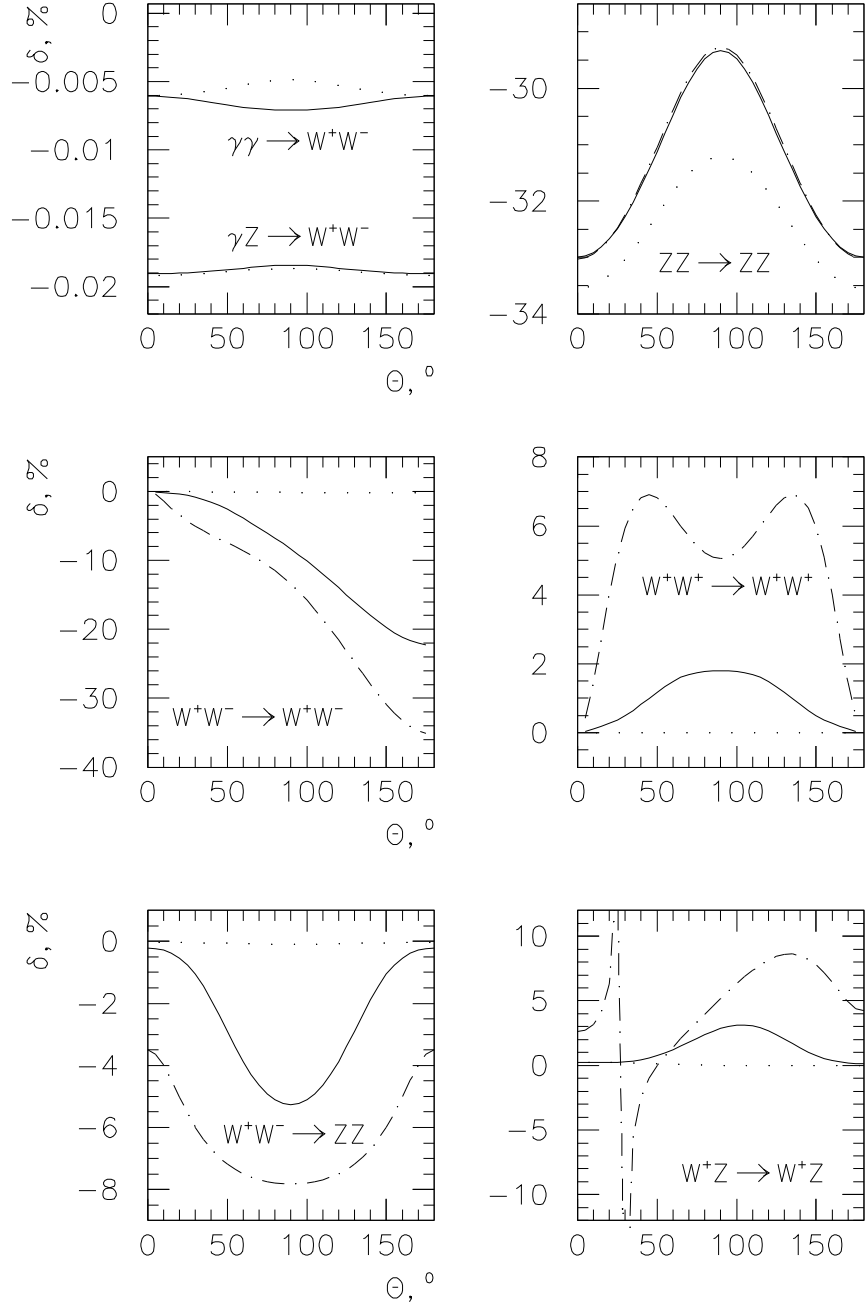,height=20cm}}
\end{picture}
\caption{Relative corrections to the Born cross sections in
Fig.~3. Solid line denotes correction for unpolarized case $UUUU$,
dash-dotted for the longitudinal $LLLL$ polarization, and dotted line
for the transverse $TTTT$ polarization.}
\end{figure}

\begin{table}[p]
\caption{Differential $d\sigma/d\cos(\theta)$ and total Born cross
sections and relative corrections for unpolarized, purely longitudinal
$LLLL$ and purely transverse $TTTT$ vector boson scattering at the
center-of-mass energy $\sqrt{s}=500$~GeV and $m_H=900$~GeV.}
\begin{center}
\begin{tabular}{|c|c|c|c|c|c|c|}\hline
 $\gamma\gamma\to W^+W^-$&$\sigma^{unpol},pb$&$\delta^{unpol},\%$ 
&$\sigma^{LLLL},pb$&$\delta^{LLLL},\%$&$\sigma^{TTTT},pb$&$\delta^{TTTT},
\%$\\\hline
$30^\circ$&61.97&$-$6.34$\cdot 10^{-3}$&---&---&60.94&$-$5.84$\cdot 10^{-3}$\\ 
$60^\circ$&8.598&$-$6.87$\cdot 10^{-3}$&---&---&7.995&$-$5.28$\cdot 10^{-3}$\\
$90^\circ$&4.551&$-$7.09$\cdot 10^{-3}$&---&---&4.044&$-$4.86$\cdot 10^{-3}$\\
$0^\circ<\theta<180^\circ$
&77.48&$-$6.30$\cdot 10^{-3}$&---&---&76.10&$-$5.84$\cdot 10^{-3}$\\
\hline\hline
 $\gamma Z\to W^+W^-$&$\sigma^{unpol},pb$&$\delta^{unpol},\%$ 
&$\sigma^{LLLL},pb$&$\delta^{LLLL},\%$&$\sigma^{TTTT},pb$&$\delta^{TTTT},
\%$\\\hline
$30^\circ$&171.0&$-$1.90$\cdot 10^{-2}$&---&---&219.4&$-$1.91$\cdot 10^{-2}$\\
$60^\circ$&22.66&$-$1.87$\cdot 10^{-2}$&---&---&28.76&$-$1.88$\cdot 10^{-2}$\\
$90^\circ$&11.48&$-$1.84$\cdot 10^{-2}$&---&---&14.54&$-$1.87$\cdot 10^{-2}$\\
$0^\circ<\theta<180^\circ$
&213.1&$-$1.90$\cdot 10^{-2}$&---&---&274.0&$-$1.91$\cdot 10^{-2}$\\
\hline\hline
 $W^+W^-\to W^+W^-$&$\sigma^{unpol},pb$&$\delta^{unpol},\%$ 
&$\sigma^{LLLL},pb$&$\delta^{LLLL},\%$&$\sigma^{TTTT},pb$&$\delta^{TTTT},
\%$\\\hline
$30^\circ$&828.0&$-$0.832&1176.&$-$5.05&1163.&$-$4.50$\cdot 10^{-2}$\\    
$60^\circ$&108.1&$-$3.87&441.0&$-$8.58&94.56&$-$8.85$\cdot 10^{-2}$\\     
$90^\circ$&42.08&$-$8.36&243.2&$-$13.4&21.74&$-$0.131\\                   
$120^\circ$&23.90&$-$14.0&144.1&$-$21.6&11.85&$-$0.157\\                   
$150^\circ$&16.73&$-$19.7&96.91&$-$30.9&10.57&$-$0.178\\                   
$10^\circ<\theta<170^\circ$
&572.1&$-$1.46&939.7&$-$8.01&731.0&$-$3.82$\cdot 10^{-2}$\\ 
\hline\hline
 $W^+W^+\to W^+W^+$&$\sigma^{unpol},pb$&$\delta^{unpol},\%$ 
&$\sigma^{LLLL},pb$&$\delta^{LLLL},\%$&$\sigma^{TTTT},pb$&$\delta^{TTTT},
\%$\\\hline
$30^\circ$&435.4&0.496&327.9&5.94&695.9&$-$5.71$\cdot 10^{-3}$\\ 
$60^\circ$&61.07&1.49&130.9&6.31&86.13&$+$6.02$\cdot 10^{-3}$\\     
$90^\circ$&35.44&1.79&115.7&5.05&43.15&$+$3.30$\cdot 10^{-3}$\\     
$10^\circ<\theta<170^\circ$
&573.8&0.456&526.4&4.54&845.2&$-$6.40$\cdot 10^{-3}$\\ 
\hline\hline
 $W^+W^-\to ZZ$&$\sigma^{unpol},pb$&$\delta^{unpol},\%$ 
&$\sigma^{LLLL},pb$&$\delta^{LLLL},\%$&$\sigma^{TTTT},pb$&$\delta^{TTTT},
\%$\\\hline
$30^\circ$&271.2&$-$1.06&412.6&$-$5.99&390.7&$-$4.50$\cdot 10^{-2}$\\
$60^\circ$&57.85&$-$3.87&262.7&$-$7.49&51.47&$-$7.31$\cdot 10^{-2}$\\
$90^\circ$&40.75&$-$5.26&241.1&$-$7.82&25.83&$-$9.55$\cdot 10^{-2}$\\
$0^\circ<\theta<180^\circ$
&347.1&$-$1.43&643.5&$-$6.66&477.9&$-$4.53$\cdot 10^{-2}$\\
\hline\hline
 $W^+Z\to W^+Z$&$\sigma^{unpol},pb$&$\delta^{unpol},\%$ 
&$\sigma^{LLLL},pb$&$\delta^{LLLL},\%$&$\sigma^{TTTT},pb$&$\delta^{TTTT},
\%$\\\hline
$30^\circ$&4.417&0.275&1.79$\cdot 10^{-2}$&$-$22.4&6.256&0.204\\  
$60^\circ$&7.506&1.21&10.13&1.43&7.048&0.108\\                  
$90^\circ$&16.76&2.85&51.19&5.17&13.05&$+$1.63$\cdot 10^{-2}$\\    
$120^\circ$&53.96&2.68&136.1&8.15&56.89&$-$1.07$\cdot 10^{-2}$\\ 
$150^\circ$&445.1&0.765&388.3&7.77&657.2&$-$2.16$\cdot 10^{-2}$\\
$0^\circ<\theta<180^\circ$
&281.0&0.718&239.2&6.81&414.7&$-$2.17$\cdot 10^{-2}$\\
\hline\hline
 $ZZ\to ZZ$&$\sigma^{unpol},pb$&$\delta^{unpol},\%$ 
&$\sigma^{LLLL},pb$&$\delta^{LLLL},\%$&$\sigma^{TTTT},pb$&$\delta^{TTTT},
\%$\\\hline
$30^\circ$&4.570&$-$32.3&40.41&$-$32.2&1.41$\cdot 10^{-3}$&$-$33.0\\
$60^\circ$&3.758&$-$30.5&32.97&$-$30.4&1.08$\cdot 10^{-3}$&$-$31.8\\
$90^\circ$&3.372&$-$29.3&29.56&$-$29.3&9.47$\cdot 10^{-4}$&$-$31.2\\
$0^\circ<\theta<180^\circ$
&7.799&$-$30.8&68.60&$-$30.8&2.29$\cdot 10^{-3}$&$-$32.2\\\hline
\end{tabular}
\end{center}
\end{table}

\begin{figure}[p]
\setlength{\unitlength}{1cm}
\begin{picture}(14,20)
\put(2,0){\epsfig{file=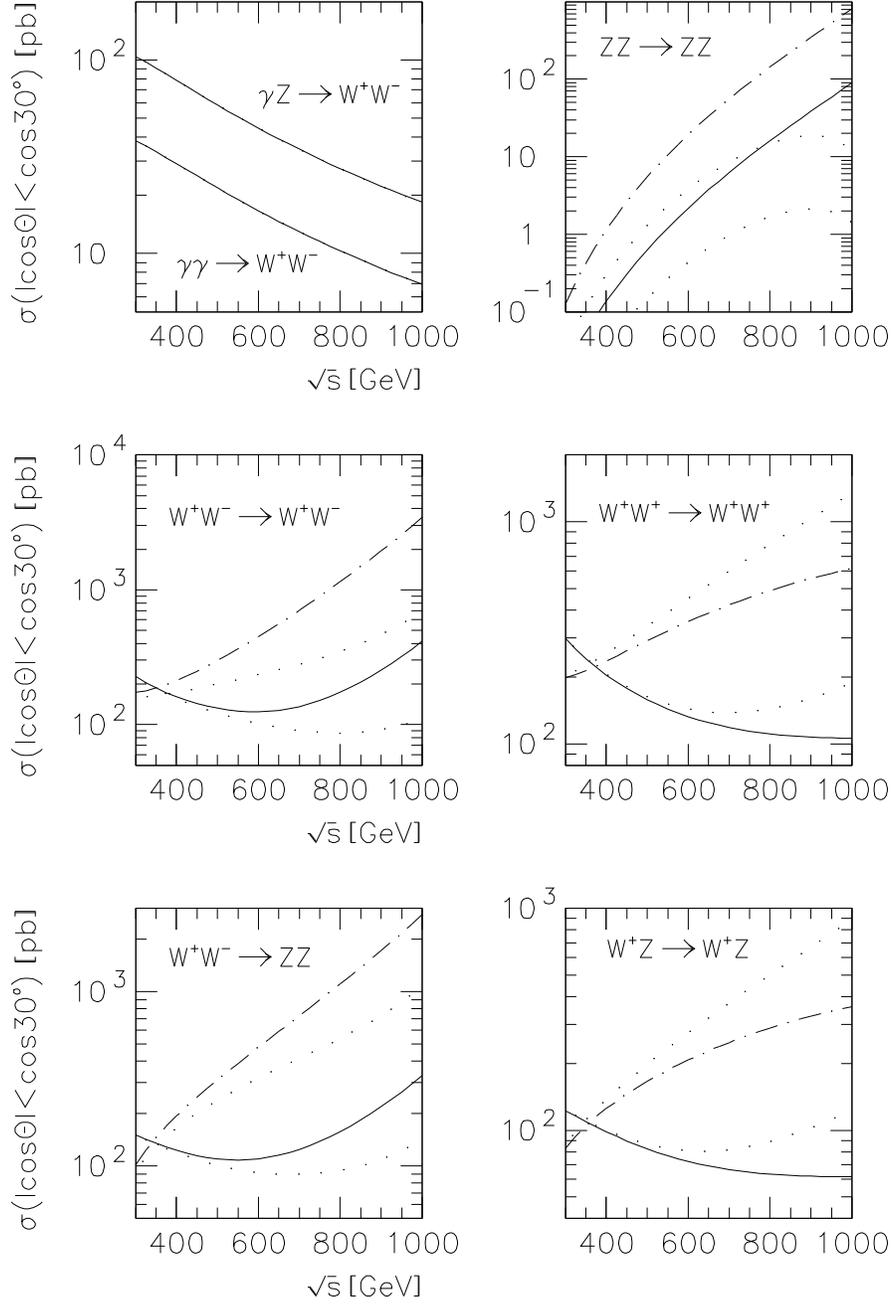,height=20cm}}
\end{picture}
\caption{Energy dependence of the cross sections for unpolarized
$UUUU$ (solid lines) and longitudinal $LLLL$ (dash-dotted lines)
vector boson scattering reactions. Dotted lines show corresponding
corrected cross sections. Higgs mass is taken to be 1.5~TeV}
\end{figure}

Relative corrections to the Born cross sections 
\begin{equation}
\delta = \frac{\sigma^{corr}-\sigma^{Born}}{\sigma^{Born}}
\end{equation}
are shown in Fig.~4.  As one can see the corrections to the
$\gamma\gamma\to W^+W^-$, $\gamma Z\to W^+W^-$ reactions are at the
level of $10^{-2}\%$ and so are unobservably small. Corrections to the
other reactions are of the order of $(10-50)\%$, {\it i.e.} quite
large, especially in some regions of phase space. Corrections for
longitudinal polarization $LLLL$ dominate, while corrections for
transverse polarization $TTTT$ are negligible for all reactions except
$ZZ\to ZZ$. Corrections for the neutral channel reactions are
negative, while for the charged channels $W^+W^+\to W^+W^+$, $W^+Z\to
W^+Z$ they are positive. The correction to the reaction $W^+Z\to W^+Z$ for
$LLLL$ polarization exhibits a pole behavior at $\theta\approx
28^\circ$, due to the fact that longitudinal Born cross section is
exactly equal to zero at this value of $\theta$.

Differential cross sections at several angles, integrated cross
sections and relative corrections are presented in Table~1. Since
complete one-loop \SM{} electroweak corrections are known for the
reactions $\gamma\gamma\to W^+W^-$ \cite{AAWW}, $ZZ\to ZZ$ \cite{ZZZZ}
and $W^+W^-\to W^+W^-$ \cite{WWWW}, one can compare our two-loop
$m_H^2$ corrections with these one-loop results. 

One-loop electroweak corrections to the cross section of the
$W^+W^-$-pair production in photon-photon collisions are about
$-(1- 3)\%$ at 500~GeV, so that there is no chance to separate the
$m_H^2$ effects from all other corrections in this reaction.

One-loop corrections to the $W^+W^-\to W^+W^-$ and $ZZ\to ZZ$
scattering reactions for longitudinal $LLLL$ polarization at 500~GeV
are positive and of the order of 10\% and $(30- 40)\%$,
respectively. So the two-loop large Higgs mass corrections
proportional to $m_H^2$ for these reactions are approximately the same
in size as one-loop electroweak corrections, but opposite in sign,
almost canceling out in the complete amplitude. It means that in order
to experimentally extract physically interesting two-loop $m_H^2$
quantum corrections from the cross sections of vector boson scattering
reactions one should necessarily include all one-loop electroweak
radiative corrections. Since enhanced two-loop corrections are
approximately of the same size as one-loop corrections, experimental
accuracy which is enough to measure the effects of one-loop
corrections is simultaneously enough to measure the contribution of
the large Higgs mass two-loop corrections.

Finally, in order to demonstrate the potential importance of large
Higgs mass corrections at high energies, we present in Fig.~5 the
energy dependence of the Born and corrected cross section of vector
boson scattering integrated over scattering angles in the region
$30^\circ < \theta < 150^\circ$ for $\sqrt{s_{VV}}$ up to 1~TeV for
the very heavy Higgs boson mass of 1.5~TeV. The existence of a
physical Higgs particle with such large mass seems to be excluded due
to triviality bounds (see \cite{LP'97-2} and references therein), and
the same conclusion follows from the non-perturbative $1/N$ approach
with the account of next-to-leading corrections \cite{1/N}. We can
consider however such a value of the $m_H$ as an effective ultraviolet
cut-off in the theory without visible scalar Higgs particle. We see
that the growth with energy of the longitudinal vector boson
scattering cross sections, which is the experimental indication of the
existence of heavy Higgs sector and/or strong interactions among
longitudinal $W_L$, $Z_L$ bosons, is strongly modified by the two-loop
$m_H^2$ corrections.  Again at high energy the cross sections of
neutral channel reactions are diminished, and those of charged channel
reactions are enhanced.  Of course at center-of-mass energy of 1~TeV
$s_{VV}$ is not very much smaller than $m_H^2$, which is the condition
under which our low-energy effective action was
calculated. Nevertheless, we think that the qualitative trend, namely
the fact that the account of large Higgs mass corrections at high
energy can change the value of the cross section by a large factor of
$2- 4$, is important for all considerations of the signal from
strong scattering of longitudinal vector boson at TeV energy.

In fact using the results of a thorough phenomenological analysis of the
effects of anomalous quartic couplings in $e^\pm e^-$ collisions
\cite{ee-nnVV,ee-VVV} we can estimate the potential of TeV $e^\pm e^-$
linear colliders in investigating the effects of enhanced $m_H^2$
two-loop corrections more quantitatively. Anomalous quartic couplings
are defined in Ref. \cite{ee-nnVV} through the following effective
electroweak chiral Lagrangians:
\begin{eqnarray}
{\cal L}_4 &=& g^4\alpha_4\Biggl[\frac{1}{2}[(W^+W^-)^2+(W^{+2})(W^{-2})]
+\frac{1}{\cw^2}(W^+Z)(W^-Z)+\frac{1}{4\cw^4}Z^4\Biggr],\label{L4}\\
{\cal L}_5 &=& g^4\alpha_5\Biggl[(W^+W^-)^2
+\frac{1}{\cw^2}(W^+W^-)Z^2+\frac{1}{4\cw^4}Z^4\Biggr],\label{L5}
\end{eqnarray}
where $g=e/\sw$. These operators introduce all possible quartic
couplings among the weak gauge bosons, that are compatible with
custodial $SU(2)_c$ symmetry \cite{chiral}. Although our complete
effective action given in Section~4 does not obey this symmetry and as
a consequence can not be described by the combination of operators
(\ref{L4}), (\ref{L5}), the dominating terms which originate from
two-loop Higgs self energy insertions in the Higgs reducible graphs
have exactly the structure of Lagrangian (\ref{L5}). Using our
expressions (\ref{WWWW}), (\ref{WWZZ}), (\ref{ZZZZ}) we can {\it
calculate} the coupling constant $\alpha_5$:
\begin{equation}
\alpha_5 \approx - \frac{g^2}{(16\pi^2)^2}\frac{m_H^2}{\mW^2}.
\label{a5}
\end{equation}
In our approach the constant $\alpha_4$ should be about an order of
magnitude smaller. The 90\% bound, based on the hypothesis
$\alpha_5=0$, obtained by combining the $e^+e^-\to\nu_e\bar\nu_e
W^+W^-$ and $e^+e^-\to\nu_e\bar\nu_e ZZ$ channels is $|\alpha_5|\le
1.5\times 10^{-3}$ for $\sqrt{s}=1.6$~TeV and integrated luminosity of
500~fb$^{-1}$ \cite{ee-nnVV}. For the Higgs mass of 1.5~TeV the value
of $\alpha_5$ from Eq. (\ref{a5}) is approximately $-6\times 10^{-3}$
(and $-2\times 10^{-3}$ for $m_H=900$~GeV), which is four times larger
than the achievable experimental limit. This comparison is a very good
indication that in the case, if a heavy Higgs scenario of the
electroweak symmetry breaking is realized in nature, its indirect
quantum effects could be measured.

\section{Conclusions}

Owing to the enhanced sensitivity to the heavy Higgs boson sector of
the \SM{}, two-loop large Higgs mass quantum corrections to low energy
vertices growing like $m_H^2$ found continuous interest in the
literature, where corrections to weak vector boson propagators
\cite{rho,masses,pisa}, triple vector boson vertices \cite{triple}, as
well as corrections due to an arbitrary gauge-invariant
non-renormalizable potential of a heavy Higgs particle \cite{hhh} were
calculated, but found to be very small and beyond experimental
verification. We have completed the existing results by calculating
the two-loop $m_H^2$ correction for $m_H\gg \mW$ to quartic
electroweak boson vertices at low energy, which were the last
quantities, which exhibit power $m_H^2$ enhancement at the two-loop
level.

Corrections are found to be especially large for the $W^+W^-W^+W^-$,
$W^+W^-ZZ$ and $ZZZZ$ quartic vertices, which receive contribution
from two-loop Higgs self-energy graphs. The value of these corrections
for vector boson scattering reactions at $\sqrt{s}=500$~GeV is found
to be of the order of $(5-30)\%$ for cross sections of the
longitudinally polarized particles, which are the most sensitive to
the $m_H^2$ effects.

By comparison with the results of phenomenological analysis of the
effects of anomalous quartic couplings in the $WW$, $ZZ$-fusion
reactions in $e^+e^-$ collisions \cite{ee-nnVV} we found that the
anomalous interactions generated by the two-loop large Higgs mass
quantum corrections seem to be large enough to be observable at the TeV
energy colliders.

An important point is the validity of the perturbation theory for
large values of Higgs mass and self-couplings. The physical Higgs
boson seems to be excluded by purely theoretical reasons at $m_H$
above 1~TeV \cite{LP'97-2}. However, it is quite encouraging that for
the Higgs mass below 1~TeV the perturbative two-loop corrections, that
are substantial when compared to the one-loop and tree level, turn out
to be remarkably close to the nonperturbative results obtained in
next-to-leading approximation in $1/N$ expansion \cite{1/N}.

\section*{Acknowledgements}

G.J. is indebted to J.J.~van~der~Bij, G.~Degrassi and A.~Ghinculov for
very useful discussions. This work was supported in part by the
Alexander von Humboldt Foundation and the Russian Foundation for Basic
Research grant 96-02-19-464.


\begin{thebibliography}{**}

\bibitem{LP'97-1} J.~Timmermans, talk given at the {\it XVIII
International Symposium on Lepton and Photon Interactions}, 28 July
1997 -- 1 August 1997, Hamburg.

\bibitem{LP'97-2} G.~Altarelli, {\it ibid.}.

\bibitem{ALR}
P. Langacker, talk at International Workshop on Supersymmetry and 
Unification of Fundamental Interactions (SUSY 95), Palaiseau, France, 
15-19 May 1995, NSF-ITP-95-140, October 1995,  hep-ph/9511207;\\
W. Hollik, talk at 11th Topical
Workshop on Proton-Antiproton Collider Physics (PBARP 96), Padua, 
Abano Terme, Italy, 26~May - 1~June 1996,
Karlsruhe University preprint No KA-TP-19-1996, hep-ph/9608325;\\
G.~Passarino, talk at CRAD96, Cracow, August 1996, hep-ph/9604344;\\
U.~Baur and M.~Demarteau, in: {\it Proceedings of the DPF/DPB Summer Study},
Snowmass 1996.

\bibitem{800}
G.~Degrassi, P.~Gambino, M.~Passera, and A.~Sirlin, CERN-TH-97-197,
hep-ph/9708311, August 1997.

\bibitem{scattering}
L.~Durand, P.N.~Maher and K.~Riesselmann, \PR D48 1993 1084;\\
K.~Riesselmann, \PR D53 1996 6226;\\
K.~Riesselmann and S. Willenbrock, \PR D55 1997 311.

\bibitem{h-numeric}
A.~Ghinculov and  J.J.~van~der~Bij, \NP B436 1995 30.

\bibitem{MDR}
P.N.~Maher, L.~Durand and K.~Riesselmann, \PR D48 1993 1061 ;
\E D52 1995 553 .

\bibitem{h-analytic}
V.~Borodulin, G.~Jikia, \PL B391 1997 434.

\bibitem{higgs-pole}
A.~Ghinculov, T.~Binoth, \PL B394 1997 139.

\bibitem{fermi_G}
A.~Ghinculov, \PL B337 1994 137;{~} \E B346 1995 426 .

\bibitem{fermi_DKR}
L.~Durand, B.A.~Kniehl and K.~Riesselmann, \PRL 72 1994 2534;{~}
\E 74 1995 1699 ; \PR D51 1995 5007.

\bibitem{vector_G}
A.~Ghinculov,  \NP B455 1995 21.

\bibitem{vector_FKKR}
A.~Frink, B.A.~Kniehl, D.~Kreimer and K.~Riesselmann,  \PR D54 1996 4548.

\bibitem{LHC}
A.~Ghinculov and J.J. van der Bij, \NP B482 1996 59.

\bibitem{mumu}
T.~Binoth, A.~Ghinculov, \PR D56 1997 3147.

\bibitem{1/N} A.~Ghinculov, T.~Binoth, J.J. van der Bij,
FREIBURG-THEP-97-19, August 1997, hep-ph/9709211; FREIBURG-THEP-97-20,
September 1997, hep-ph/9711318.

\bibitem{screening}
M.~Veltman, {\it Acta Phys. Pol.} {\bf B8} (1977) 475;
\PL B70 1977 253;{} \quad \PL B91 1980 95.

\bibitem{rho}
J. van der Bij, M. Veltman, \NP B231 1984 205.

\bibitem{masses}
J.J. van der Bij, \NP B248 1984 141 .

\bibitem{pisa}
R. Barbieri, P. Ciafaloni, A. Strumia, \PL B317 1993 381.

\bibitem{proof}
M.B.~Einhorn, J.~Wudka, \PR D39 1989 2758;{}
\PR D47 1993 5029.

\bibitem{triple}
J.J. van der Bij, \NP B255 1985 648.

\bibitem{OS}
K.-i.~Aoki, Z.~Hioki, R.~Kawabe, M.~Konuma and T.~Muta,
Suppl. {\it Prog. Theor. Phys.} {\bf 73} (1982) 1;\\
M.~B\"ohm, H.~Spiesberger and W.~Hollik, 
{\it Fortschr. Phys.} {\bf 34} (1986) 687.

\bibitem{pole}
S.~Willenbrock and G.~Valencia, \PL B247 1990 341.

\bibitem{stuart} 
R.G.~Stuart, \PL B272 1991 353.

\bibitem{Tkachov}
F.V.~Tkachov, \PL B124 1983 212 ; \IJMP A8 1993 2047.

\bibitem{parts}
F.V.~Tkachov, \PL B100 1981 65;{}\\
K.G.~Chetyrkin, F.V. Tkachov, \NP B192 1981 159.

\bibitem{ET}
B.W.~Lee, C.~Quigg, and H.~Thacker, \PR D16 1977 1519;{}\\
M.S.~Chanowitz and M.K.~Gaillard, \NP B261 1985 379;{}\\
G.J.~Gounaris, R.~K\"ogerler and H.~Neufeld, \PR D34 1986 3257.

\bibitem{chiral} 
T.~Appelquist and C.~Bernard, \PR D22 1980 200;{}\\
A.~Longhitano, \NP B188 1981 118;{}\\ 
A.~Falk, M.~Luke and E.H.~Simmons, \NP B365 1991 523;{}\\ 
F.~Feruglio, \IJMP A8 1993 4937;{}\\
T.~Appelquist and G.H.~Wu, \PR D48 1993 3235;{}\\
F.~Boudjema, in Proceedings of the Workshop {\it $e^+e^-$
Collisions at 500~GeV: the Physics Potential, Part~C},
Munich--Annecy--Hamburg, 1993, DESY 93-123C, p. 177.

\bibitem{ee-nnVV} 
E.~Boos, H.J.~He, W.~Kilian, A.~Pukhov, C.P.~Yuan, P.M.~Zerwas,
DESY-96-256, August 1997, hep-ph/9708310;\\
T.~Han, H.-J.~He, C.P.~Yuan, UCD-97-22, November 1997, hep-ph/9711429.

\bibitem{ee-VVV}
A.~Ghinculov, J.J. van der Bij \PL B279 1992 189;{}\\
G.~B\'elanger and F.~Boudjema, \PL  B288 1992 210;{}\\
S. Dawson, A. Likhoded, G. Valencia, O. Yushchenko, Proceedings of 1996
DPF / DPB Summer Study on New Directions for High-Energy Physics,
Snowmass, CO, 25 June - 12 July 1996,  hep-ph/9610299;\\
O.J.P.~\'Eboli, M.C.~Gonzalez-Garcia, and J.K.~Mizukoshi,
IFT-P.079/97, hep-ph/9711499.

\bibitem{AAWW}
A.~Denner, S.~Dittmaier, and R.~Schuster, \NP B452 1995 80;{}\\
G.~Jikia, \NP B494 1997 19.

\bibitem{ZZZZ}
A.~Denner, S.~Dittmaier, T.~Hahn, \PR D56 1997 117.

\bibitem{WWWW}
A.~Denner, T.~Hahn, PSI-PR-97-31, November 1997, hep-ph/9711302.

\bibitem{hhh}
J.J. van der Bij, \NP B267 1986 557.

\end{thebibliography}
\end{document}